\documentstyle[12pt,epsfig]{article}

\newcommand{\beq}{\begin{equation}}
\newcommand{\eeq}{\end{equation}}
\newcommand{\be}{\begin{eqnarray}}
\newcommand{\ee}{\end{eqnarray}}

\def\nue{{\nu_e}}

\newcommand{\dm}{\mbox{$\Delta{m}^{2}$~}}

\def\kl{{KamLAND~}}

\begin{document}
\title{Testing the solar LMA region with KamLAND data}

\author
{Abhijit Bandyopadhyay$^a$\thanks{e-mail: abhi@theory.saha.ernet.in},
Sandhya Choubey$^b$\thanks{e-mail: sandhya@he.sissa.it},
Srubabati Goswami$^c$\thanks{e-mail: sruba@mri.ernet.in},\\
Raj Gandhi$^c$\thanks{e-mail: raj@mri.ernet.in},
D.P.Roy$^d$\thanks{e-mail: dproy@theory.tifr.res.in}
\\
\\
$^a$Saha Institute of Nuclear Physics,\\
1/AF, Bidhannagar, Calcutta 700 064, INDIA.\\
\\
$^b$ Scuola Internazionale Superiore di Studi Avanzati\\
I-34014, Trieste, Italy\\
\\
$^c$Harish-Chandra Research Institute,\\
 Chhatnag Road, Jhunsi,Allahabad 211 019, India\\
\\
$^d$Tata Institute of Fundamental Research,\\
Homi Bhabha Road, Mumbai 400005, India
}
\maketitle

\begin{abstract}

 We  investigate the potential of 3 kiloTon-years(kTy)
of  
KamLAND data to further constrain the  
$\Delta m^2$ and $\tan^2\theta$ values
compared to those presently allowed by existing  KamLAND and global solar 
data. 
We study the extent, dependence and characteristics  of this 
sensitivity in 
and around the two parts of the LMA region that are currently 
allowed.  
Our analysis with  3 kTy simulated spectra shows that 
KamLAND spectrum data by itself can constrain  
$\Delta m^2$ with high precision.
Combining the spectrum with global solar   data further tightens the 
constraints on allowed values of $\tan^2\theta$ and 
$\Delta m^2$. 
We also study the effects of future neutral current data
with a total error of 7\%
from the Sudbury Neutrino Observatory. 
We find that these future measurements offer the potential of  considerable precision in determining  
the oscillation parameters (specially the mass parameter). 

\end{abstract}

\newpage

\section{Introduction}
\label{sec:introduction}

The Sudbury Neutrino Observatory (SNO) charged current (CC) and 
neutral current (NC) data on the measurement of 
solar neutrino flux  has provided strong
 evidence for neutrino flavor conversion 
\cite{Ahmad:2002jz,Ahmad:2002ka} and
it establishes the presence of $\nu_\mu/\nu_\tau$ flavor in the solar 
$\nu_e$ flux at 5.5$\sigma$ level. 
The SNO and SK results together with the data from the 
radiochemical experiments Cl \cite{Cleveland:nv} and Ga \cite{globalsolar}
{\footnote{
For recent reviews on solar neutrino experiments see
\cite{Goswami:2003b,Miramonti:wz}.}}
single out the LMA solution based on  MSW resonant matter 
conversion 
as the most probable solution of the solar neutrino puzzle
\cite{a1}-\cite{a9}. 
In \cite{Choubey:2002nc}, 
for instance, this solution is characterized by best fit values of
$\tan^2\theta= 0.41,  
\Delta m^2= 6.06 \times 10^{-5} $eV$^2$ for the neutrino mixing parameters. 
Spectacular confirmation in favour of this solution, 
using terrestrial 
neutrino sources, has come recently from the 
KamLAND experiment in Japan \cite{kldata}. 
The uniqueness of KamLAND \cite{kam_2,prekl,mur}    
lies in its sensitivity to 
masses lying 
in the LMA region through its measurement of the reactor antineutrino 
energy spectrum.   
With 162 ton-years of data, KamLAND has already split the 
allowed LMA zone into two smaller sub-zones, 
the low-LMA with $\dm \sim 7.2\times 10^{-5}$ eV$^2$ and the 
high-LMA zone with $\dm \sim 1.5\times 10^{-5}$ eV$^2$ 
\cite{kl1,kl2}.

In this paper, we consider  a broader time-frame than spanned by the 
first results, and obtain projected results and conclusions that should be 
forthcoming from a 3 kTy exposure. 
With present fiducial volume this correspods to KL run of 8 - 10 years time.
First we discuss the present constraints
and next we discuss the future perspectives in the light of the present 
data. 
We determine the allowed areas in the $\Delta m^2$-$\tan^2\theta$ plane 
from the  analysis of total  KamLAND rate, KamLAND spectrum and 
examine the role of both of these in constraining the allowed 
regions by themselves as well as in conjunction with 
the global solar data.
We simulate the projected KamLAND spectrum at several sample 
$\Delta m^2$ and 
$\tan^2\theta$ values, 
taken from the currently allowed regions, 
and attempt to delineate in detail
the limits of sensitivity for KamLAND over a
3 kTy
period. 
We study the dependence of the reconstituted 
parameter regions on the sample values of $\Delta m^2$ and
$\tan^2\theta$ chosen and demonstrate that for spectrum simulated 
at points inside both the low-LMA and the high-LMA region
 the accuracy of reconstruction is 
quite high, leading to an excellent precision in determination 
of the oscillation parameters, especially $\Delta m^2$.

Section 2 describes the salient features of the KamLAND 
detector, and the expected ${\bar {\nu_e}}$ reactor flux 
to which it is sensitive. In Section 3 we  discuss 
the analysis procedure and results. Section 4 summarises our conclusions.

\section{The KamLAND detector, 
the reactor flux of Electron Anti-Neutrinos and the Event Rate}

KamLAND \cite{kam_2} is a 1 kton liquid scintillator neutrino detector 
located at the earlier Kamiokande 
site in the Kamioka mine in Japan. 
It measures   the $\bar{\nu}_e$ flux from  
16 Japanese nuclear power reactors whose distances range from 
$80$ km  to $800$ km. However $\approx 79\%$
of the measured flux come  
from reactors situated at distances between 138 km to 214 km
from the detector. 
The reaction that detects the  
$\bar{\nue}$ is the inverse beta decay 
$\bar{\nu}_e+p\rightarrow e^{+}+n$. 
The positrons are annihilated to 
produce two $\gamma$ rays.  
Neutron capture in the medium also generates a delayed $\gamma$ 
signal.  
The correlation between these two records an event 
grossly free from the backgrounds. 

The neutrino spectrum from the fission of a particular isotope 
$j$ is
conveniently parameterizable \cite{mur,vo} 
(in units of MeV$^{-1}$ per fission)  as 
\begin{eqnarray}
d N_\nu^{j}/d E_\nu=\exp (a_0+a_1 E_\nu+ a_2 E_\nu^2).
\label{nuspectrum}
\end{eqnarray}
Here $j =1,2,3,4$, corresponding to the 4 isotopes  $^{235}$U, $^{239}$P,
$^{238}$U,  $^{241}$Pu which constitute the fuel. The fitted values of the 
$a_ks$ in the equation above are reproduced for completeness from \cite{mur,vo}
 in Table 1.

\begin{table}
\begin{center}
\caption{Fitted Parameters $a_k, k=0,1,2$ for the reactor neutrino spectrum. 
The last row shows the energy released for each isotope per fission. } 
\label{table: a_param}
\vskip 0.5cm
\begin{tabular}{|c|c|c|c|c|} \hline 
Isotope &  $^{235}$U & $^{239}$Pu & $^{238}$U & $^{241}$Pu \\ \hline 
$a_{0}$ &  0.870  & 0.896 & 0.976 & 0.793  \\ \hline 
$a_{1}$ &  -0.160 & -0.239 & -0.162 & -0.080 \\ \hline 
$a_{2}$ & -0.0910 & -0.0981 & -0.0790 & -0.1085 \\ \hline 
$\epsilon_j$ (MeV) & 201.7 & 205.0 & 210.0 & 212.4 
\\ \hline
\end{tabular}
\end{center}
\end{table}
 
In addition, each isotope has a characteristic  
energy released per fission,   $\epsilon_j$.
These are also  reproduced from \cite{vo} in Table 1.

Table 2 (from \cite{kam_2}) gives the distances of the 
various reactors from the Kamioka 
mine which houses KamLAND, along with the maximum thermal power 
$N^i_{max}$ (in Giga-watts) of each $i$, the reactor index, 
which runs from $1-16$. 
Also, we note that in principle, 
the power of each reactor varies over the year depending on demand, 
fuel composition, re-fuelling times etc.
This dependence is averaged over 
for our purposes, and we assume  that each reactor is in the running mode 
at maximum output 80\% of the time.
\begin{table}[t]
\begin{center}
\caption{Reactor distances and power outputs }
\label{table:r_param}
\vskip 0.5cm
\begin{tabular} {|l|c|c|} \hline
Reactor Site & Distance $d_i$ (km) &  Power $N_{max}^i$(Giga-watts)  \\ \hline
Kashiwazaki & 160 & 24.6 \\ \hline
Ohi & 180 & 13.7 \\ \hline
Takahama & 191 & 10.2 \\ \hline
Hamaoka & 213 & 10.6 \\ \hline
Tsuruga & 139 & 4.5 \\ \hline
Shiga & 88 & 1.6 \\ \hline
Mihama & 145 & 4.9 \\ \hline
Fukushima-1 & 344 & 14.2 \\ \hline
Fukushima-2 & 344 & 13.2 \\ \hline
Tokai-II & 295 & 3.3  \\ \hline
Shimane & 414 & 3.8 \\ \hline
Ikata &  561 & 6.0 \\ \hline
Genkai & 755 & 6.7 \\ \hline
Onagawa & 430 & 4.1 \\ \hline
Tomari & 784 & 3.3 \\ \hline
Sendai & 824 & 5.3 \\  \hline
\end{tabular} 
\end{center}
\end{table}
Using the above data, one may then write an expression for the spectrum
from a given reactor $i$ as,
\begin{eqnarray}
S_i = \sum_{j} \frac{f_j^i N_{max}^i (0.8)}
{\sum_{k} f_k^i \epsilon_k} d N_\nu^{j}/d E_\nu 
\end{eqnarray}
Here $f_j^i$ is the fractional abundance of isotope $j$ in reactor 
$i$ at a given time. The time dependence is again averaged over,
and for all the reactors we use for the abundance the values,
53.8\% for $^{235}$U, 32.8\% for $^{239}$Pu, 
7.8\% for $^{238}$U, and
5.6\% for $^{241}$Pu, as in \cite{kldata}.
Convenient units for $S_i$ are MeV$^{-1}$sec$^{-1}$, obtained by converting 
the power in Giga-watts into Mev per sec by multiplying by the factor $6.24 \times
10^{21}$.

The other  quantities needed to determine the event rate are the 
cross-section, the survival probability for anti-neutrinos and the number 
of free proton
targets in the scintillator. The cross-section is given by
\begin{equation}
\sigma(E_{\nu})=\frac{2 \pi^2}{m_e^5 f \tau_{n}}p_{e^+} E_{e^+}.
\end{equation}
Here $f=1.69$ is the integrated Fermi function for neutron 
$\beta$-decay, $m_{e}$ is the positron mass, $E_{e^+}$ is the 
positron energy, $p_{e^+}$ is the positron momentum and  
$\tau_{n}=886.7 $ secs is the neutron lifetime. 
The total {\it visible
energy} ($E_{vis}$) corresponds to
$E_{e^+}+m_{e}$, where $E_{e^+}$ is the total energy of the positron and
$m_e$ the electron mass. The total positron energy
is related to the incoming antineutrino energy $E_\nu$ through the
relation,
$E_{e^+}=E_{\nu}-(m_n -  m_p) - {\bar E_{rec}}$, where $m_n - m_p = 1.293$ MeV is the
neutron--proton mass difference 
and ${\bar E_{rec}}$ is the average neutron
recoil energy calculated here  using \cite{beavo}.

The two-generation 
survival probability for the antineutrinos from each of the reactors 
$i$ is given by
\begin{eqnarray}
P_i(\bar{\nu}_e\leftrightarrow\bar{\nu}_e) &= &  
 1 - \sin^22\theta\sin^2\left(\frac{1.27\Delta m^2d_i}
{E_{\nu}}\right)
\label{prob}
\end{eqnarray} 
where $d_i$ is the distance of reactor $i$ to KamLAND in $km$, 
$E_{\nu}$ is in GeV and $\Delta m^2$
is in eV$^2$. 
Thus the total observed event-rate in KamLAND is given by
$ N_{KL}$ in 
sec$^{-1}$,
\begin{eqnarray}
N_{KL} &= &  \nonumber
\eta N_p  \times \int dE_{vis}\int dE_{e^+}  R(E_{vis},E_{e^+})  \times \\ 
 & &\sigma(E_{e^+} + 1.293) 
 \sum_{i} S_i \frac{P_i(\bar{\nu}_e
\leftrightarrow\bar{\nu}_e)}{4\pi d_i^2}  
\label{eventrate}
\end{eqnarray}
where $N_p$ is the 
number of free protons in the fiducial volume of the detector.
\kl has declared 162 ton-year data corresponding to 408 tons fiducial mass 
containing 3.46$\times 10^{31}$ free protons. The integrated total power 
$P/{4\pi L^2}$ is 254 ${\rm Joule/cm^2}$. 
The efficiency ($\eta$) is 78.3\%. 
R(E$_{vis}$,E$_{e^+}$) is the energy resolution function given by
\begin{eqnarray*}
R(E_{vis},E_{e^+}) &=& \frac{1}{\sqrt{2\pi \sigma_0^2}} 
\exp \left( -\frac{(E_{vis} - E_{e^+} - m_e)^2}{2 \sigma_0^2} \right)
\end{eqnarray*}
with $\sigma_0$/E$_{e^+}$ = 7.5\%/$\sqrt{E_{e^+}}$
($E$ in MeV) \cite{kldata,panic02}.

The ratio of the observed event-rate to the expected rate 
in KamLAND is defined as 
\be
R_{KL} = \frac{N_{KL}}{N^0_{KL}}
\ee
where $N^0_{KL}$ is the expected event-rate in KamLAND given by 
Eq.(\ref{eventrate}) with $P(\bar{\nu}_e
\leftrightarrow\bar{\nu}_e)=1$.
The first \kl results report a 
$R_{Kl}=0.611 \pm 0.085 \pm 0.041$ \cite{kldata}.

\section{Analysis and Results}

\subsection{Current Data}

\begin{figure}[t]
\centerline{\psfig{figure=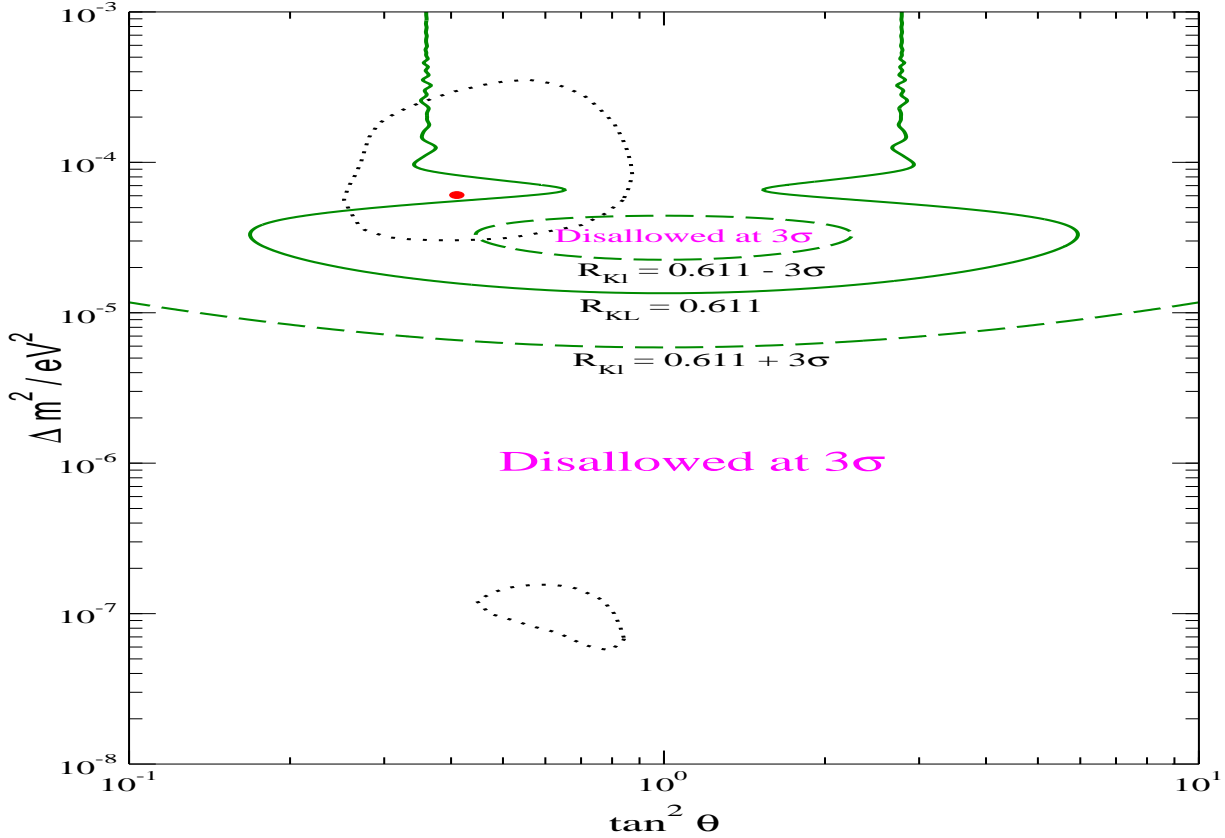,height=5.in,width=7.0in}}
\caption{The isorate lines for the KamLAND detector corresponding to
the observed rate and $\pm 3\sigma$ limits
 in the 
$\Delta m^2-\tan^2\theta$ plane. 
Also shown by dotted line is the $3\sigma$ allowed 
area from the global analysis of the solar neutrino data . The 
best-fit solution for the  solar data is marked.
}
\label{isorate}
\end{figure}

In Figure \ref{isorate} we show 
the lines of constant KamLAND rate $R_{KL}$ 
in the $\Delta m^2-\tan^2\theta$ plane for 
the observed value of 0.611 
and $\pm$ 3$\sigma$ limits. The 
best-fit point from global solar neutrino analysis and the 
corresponding $3\sigma$ contour (dashed lines) 
in the LMA and LOW regions are also shown. 
The best-fit point \cite{Bandyopadhyay:2002xj,Choubey:2002nc}
of the global solar neutrino data predicts a 
KamLAND rate of 0.65 very close to the observed rate. 
We note that a particular observed $R_{KL}$ may correspond to 
a wide range of KamLAND spectra.
The reverse, of course, is not true since an observed 
KamLAND spectrum singles out a unique rate (within errors).

We first do a statistical 
analysis with the  KamLAND rate alone. For the rate we  
define the $\chi^2$ as 
\beq
\chi^2_{KL} = \frac{(R_{KL}^{expt} - R_{KL}^{theory})^2}{\sigma^2}
\eeq
where $\sigma = \sqrt{\sigma_{syst}^2 + \sigma_{stat}^2}$, 
$\sigma_{syst}$ and $\sigma_{stat}$ being the total systematic and 
statistical error in the KamLAND data respectively. We take 6.42\%
systematic uncertainty along with the published statistical 
errors. 
To eliminate the 
geophysical background we take a visible energy 
threshold of 2.6 MeV. 
\cite{kldata,panic02}. 

For a more complete  statistical study, 
we next  do a combined analysis of global solar data  and the
observed KamLAND rate.   
We use a combined $\chi^2$ function defined as 
\beq
\chi^2 = \chi^2_{\odot} + \chi^2_{KL}
\label{chi1}
\eeq
For the $\chi^2_\odot$ (solar), we use the 
data on total rate from the Cl experiment, the
combined rate from the Ga experiments (SAGE+GALLEX+GNO),
the 1496 day data on the SK zenith angle energy spectrum and
the combined SNO day-night spectrum. We define the
$\chi^2$ function in the ``covariance'' approach as
\be
\chi^2_{\odot} = \sum_{i,j=1}^N (R_i^{\rm expt}-R_i^{\rm theory})
(\sigma_{ij}^2)^{-1}(R_j^{\rm expt}-R_j^{\rm theory})
\label{chi2}
\ee
where $R_{i}$ are the solar data points, $N$ is
the number of data points (80 in our case) and
$(\sigma_{ij}^2)^{-1}$ is the inverse of the covariance matrix,
containing the squares of the correlated and uncorrelated experimental
and theoretical errors. 
For further details of our solar analysis we refer the reader 
to \cite{Choubey:2002nc,Bandyopadhyay:2002qg}. 

\begin{figure}[t]
\centerline{\psfig{figure=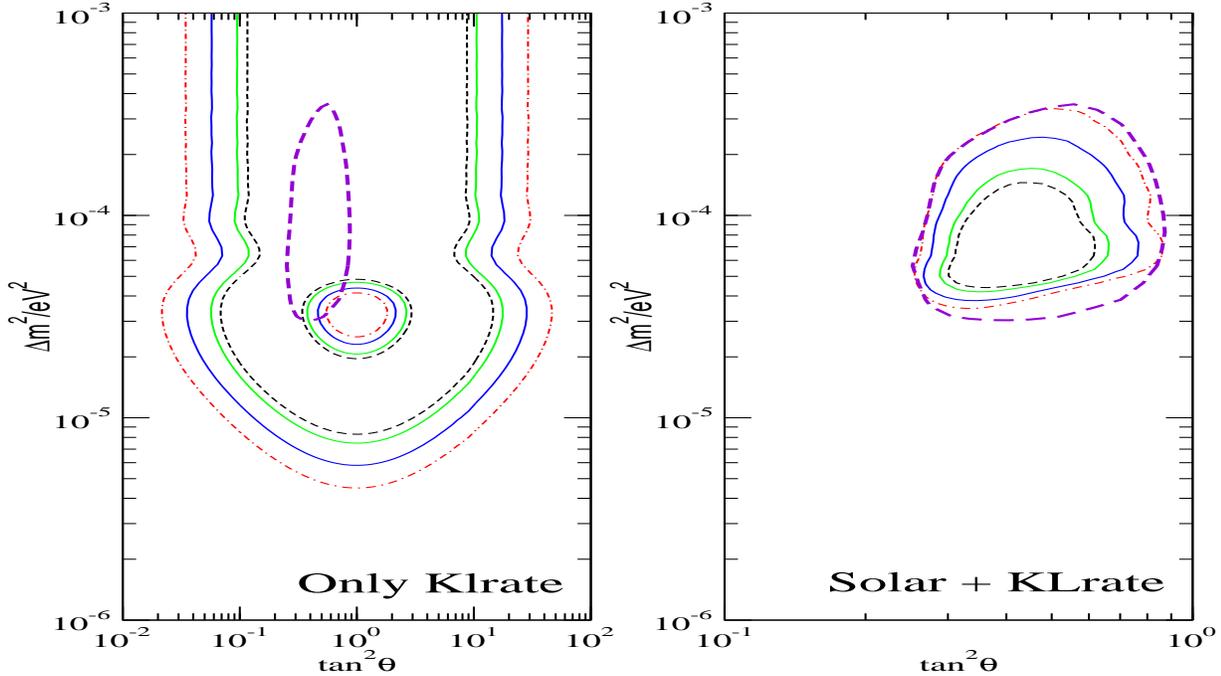,height=5.in,width=7.0in}}
\caption{The 90\%, 95\%, 99\% and 99.73\% C.L. contours 
in the LMA region. 
The left hand panel gives the allowed region from an analysis of only 
KamLAND rate. The right hand panel gives the allowed region
from a combined solar+KamLAND rate analysis.  
The purple dashed line shows the $3\sigma$ 
only solar contour.}
\label{solkamrate}
\end{figure}

In Figure \ref{solkamrate} 
we draw the 90\%, 95\%, 99\% and 99.73\% C.L. 
allowed area in the LMA region from  only \kl rate and  combined 
solar+KamLAND rate analysis. 
Superimposed on that we show the $3\sigma$ allowed area from  
solar data alone. 
The current observed \kl rate is in excellent agreement 
with the predicted \kl rate from the best-fit solution to 
the solar data. From the left-hand panel of 
Figure \ref{solkamrate} we see that the allowed areas 
from \kl rate data are mostly consistent with the $3\sigma$ allowed 
area from the global solar data, except from a small region 
around the bottom right-hand edge of the latter. This fact 
is again reflected in the global solar and \kl rate 
allowed areas shown in the right-hand panel. The inclusion 
of the \kl rate into the global analysis allows most of the 
regions allowed before, except for a small region around 
low \dm and large $\tan^2\theta$ (the bottom right zone).

\begin{figure}[t]
\centerline{\psfig{figure=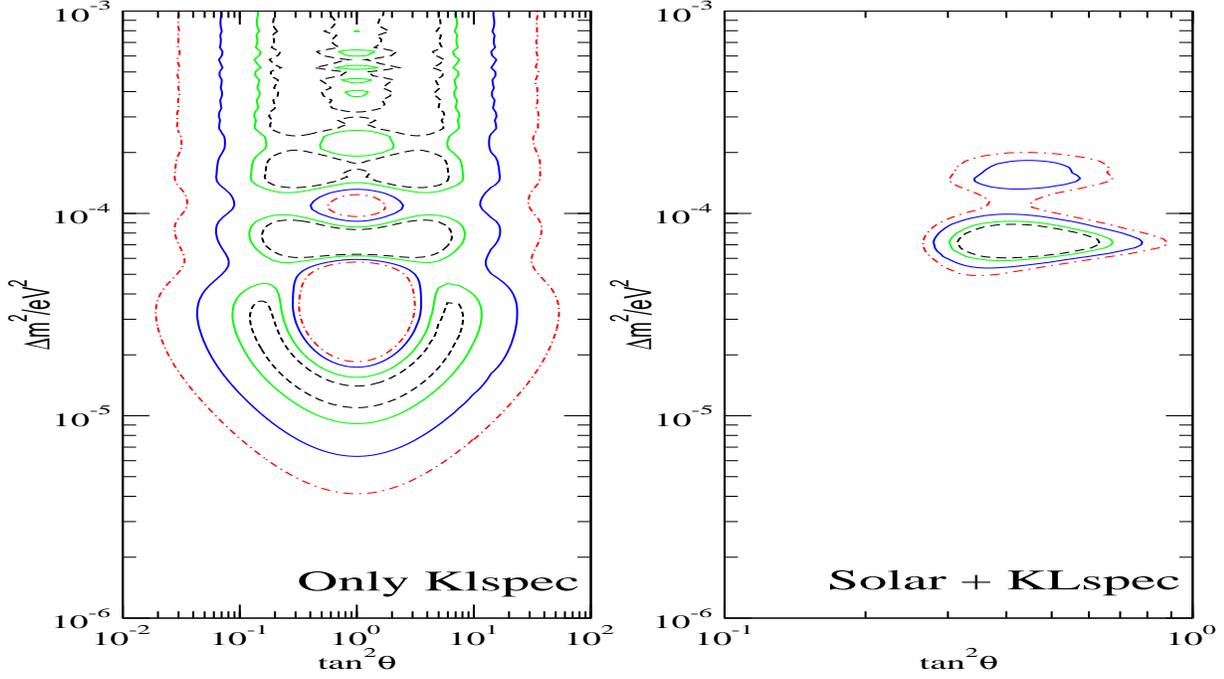,height=5.in,width=7.0in}}
\caption{The 90\%, 95\%, 99\% and 99.73\% C.L. contours 
in the LMA region. 
The left hand panel gives the allowed region from an analysis of only 
KamLAND spectrum data. The right hand panel gives the allowed region
from a combined solar+KamLAND spectrum analysis.}  
\label{solkamspecdata}
\end{figure}

Next, our aim is to see how far the allowed areas can be 
constrained with the inclusion of KamLAND   
spectrum data.  The current \kl spectrum data are 
rather low on statistics and hence we consider a 
Poisson distribution for the spectral events.  
Thus the $\chi^2$ for the current KamLAND spectrum 
is defined as 
\be
\chi^2_{KLspec}=
\sum_{i}\left[2(X_n N_i^{th} - N_i^{exp}) + 2 N_i^{exp} ln(\frac{N_i^{exp}}
{X_n N_i^{th}})\right] + \frac{(X_n -1)^2}{\sigma^2_{sys}}
\label{chip}
\ee
where $\sigma_{sys}$ is taken to be 6.42\%, $X_n$ 
is a normalization allowed 
to vary freely, and the 
sum is over the KamLAND spectral bins.

The left hand panel in Figure 
\ref{solkamspecdata} shows the allowed area from 
only \kl spectrum analysis. 
The right hand panel in Figure \ref{solkamspecdata} 
gives the allowed region
in the oscillation parameter space after including the global solar data. 
These contours are obtained by minimising  
\beq
\chi^2 = \chi^2_{\odot} + \chi^2_{KLspec}
\label{chi1spec}
\eeq
Together the \kl and solar data are 
instrumental in narrowing down 
the parameter range by a large amount. 
At 99\% C.L. the allowed LMA region is bifurcated into two parts 
-- a low-LMA zone around the best-fit point  
$\dm=7.17\times 10^{-5}$ eV$^2$ and $\tan^2\theta_{12}=0.44$
and a high-LMA zone 
around a second best-fit of  $\dm=1.49\times 10^{-4}$ eV$^2$
and $\tan^2\theta_{12}=0.43$. At 3$\sigma$ the two regions merge.  

\subsection{ 3 kton year data}
\begin{figure}[t]
\centerline{\psfig{figure=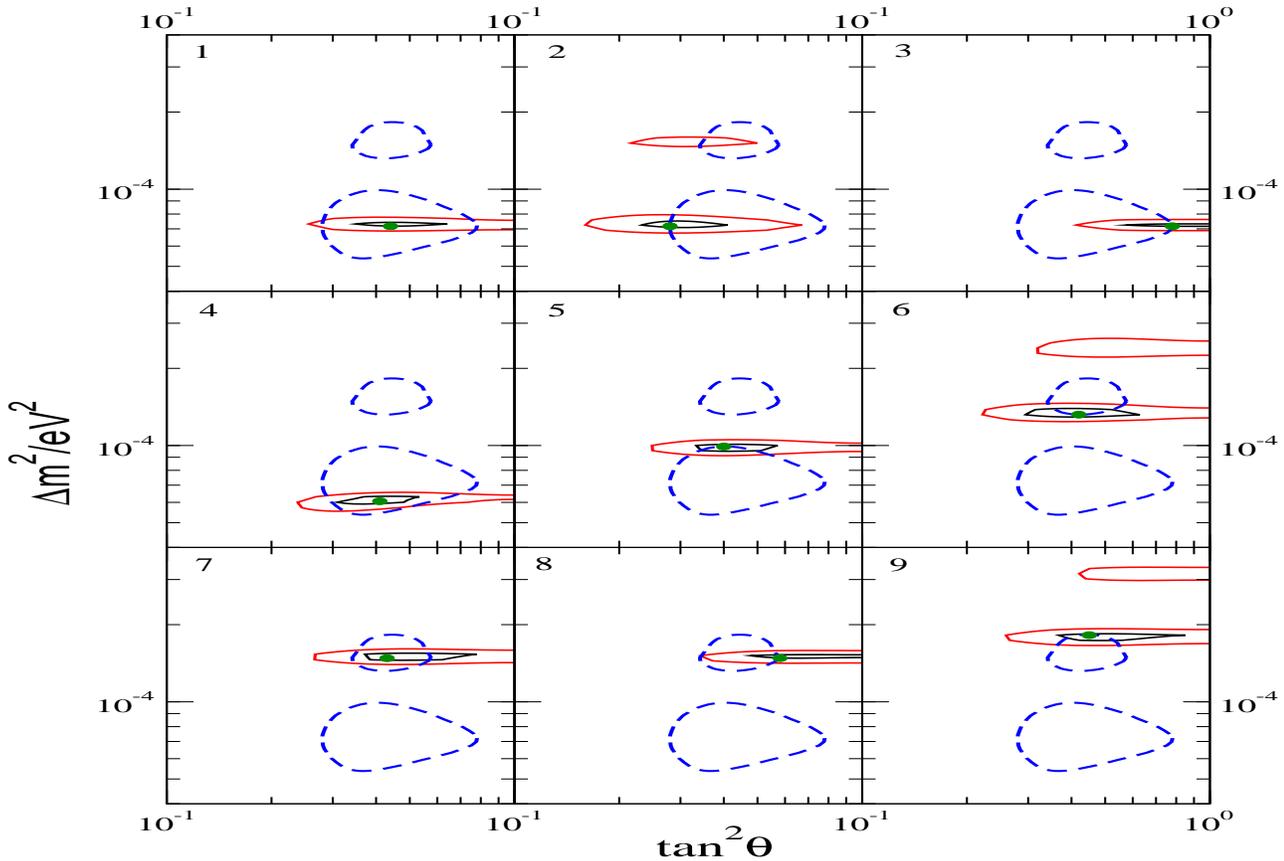,height=5.in,width=7.0in}}
\caption{The 1$\sigma$ and 3$\sigma$ contours for
the 3 kTy projected KamLAND spectrum analysis alone.
The different panels are for the simulated spectrum at
different fixed values of $\Delta m^2$ and $\tan^2\theta$ shown by
bold dots. For reference we also show by 
dashed lines the current 99\% C.L. allowed areas 
from the global analysis of the solar and \kl data.}
\label{kamspec} 
\end{figure}

\begin{figure}[t]
\centerline{\psfig{figure=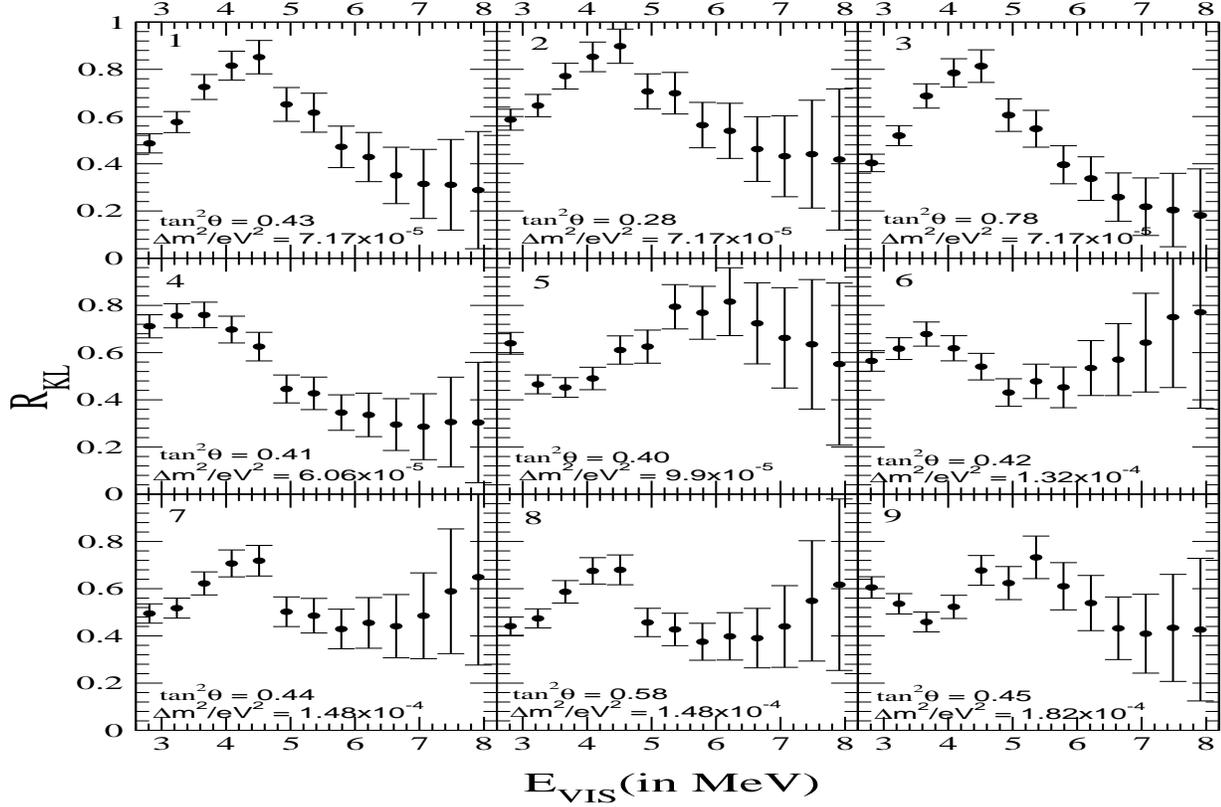,height=5.in,width=7.0in}}
\caption{Ratio of oscillated to unoscillated
events for the 3kTy simulated KamLAND spectrum for the different sets
of 
$\Delta m^2$ and $\tan^2\theta$ corresponding to Figure \ref{kamspec}.}
\label{spectrum}
\end{figure}

\begin{figure}[t]
\centerline{\psfig{figure=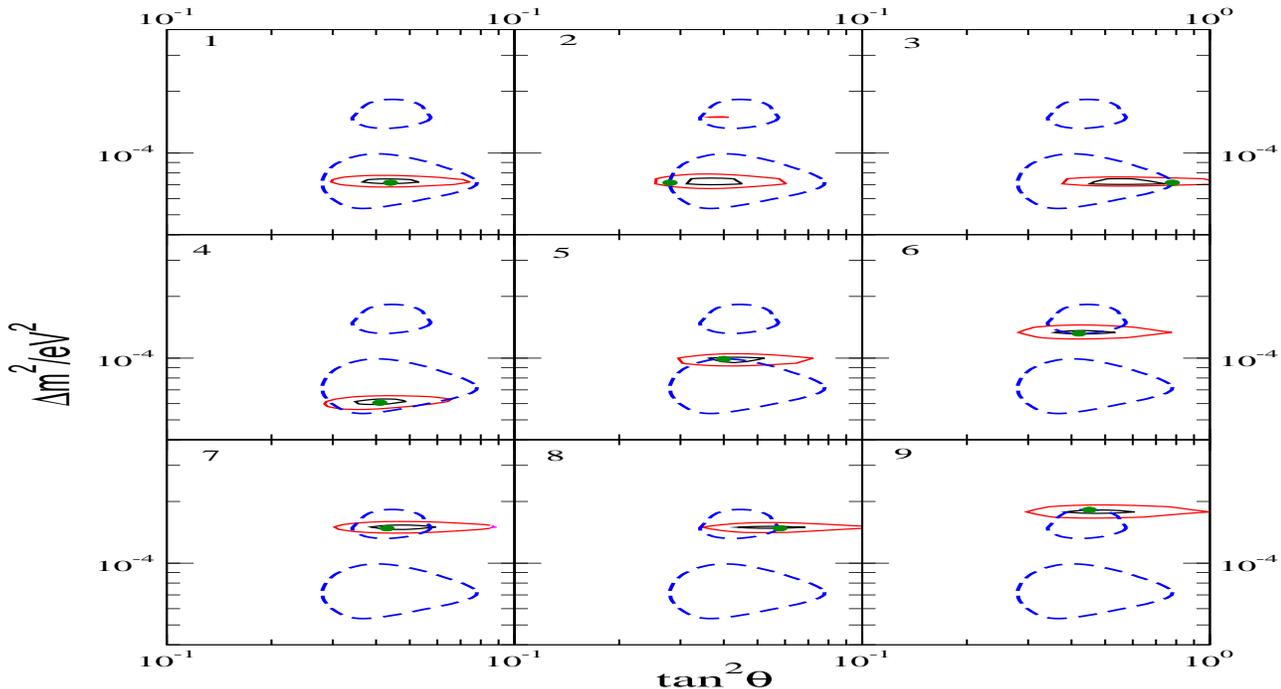,height=4.in,width=7.0in}}
\caption{The 1$\sigma$ and 3$\sigma$ contours for
the combined analysis including the global solar data,
the projected 3 kTy \kl data and the CHOOZ reactor data \cite{chooz}.
The different panels are for the simulated spectrum at
values of $\Delta m^2$ and $\tan^2\theta$ of
the corresponding panels
of Figures \ref{kamspec} and \ref{spectrum}.
The dashed line shows the presently allowed $3\sigma$
contour from solar+\kl spectrum analysis.}
\label{solkamspec}
\end{figure}

\begin{figure}[t]
\centerline{\psfig{figure=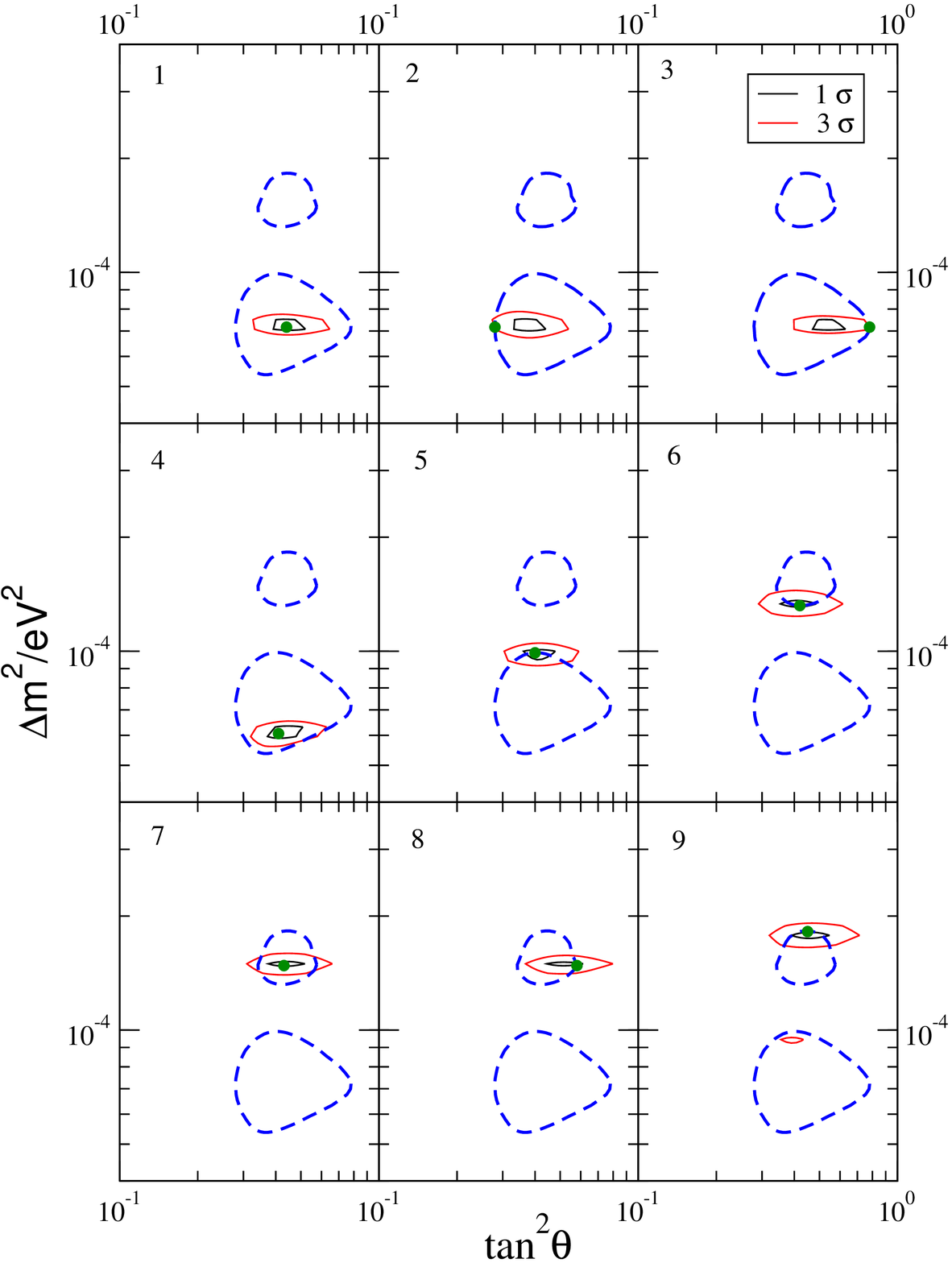,height=4.in,width=7.0in}}
\caption{
Same as in Figure \ref{solkamspec} but with a
smaller future projected total
uncertainty in the SNO NC rate of only 7\%.}
\label{sno7}
\end{figure}

In this section we investigate the evolution of the 
allowed zones as \kl collects more statistics for its spectrum data. 
We do a projected analysis
with 3 kTyr \kl spectrum simulated at few 
representative 
values of $\Delta m^2$ and $\tan^2\theta$. 
We randomize the generated spectra  
to take into account the possible fluctuations. 
We use these simulated spectra in a $\chi^2$ analysis and 
reconstruct the allowed regions in the $\Delta m^2$--$\tan^2\theta$ 
parameter space. Since statistics are expected to be 
large for the 3 kTy exposure, we consider a Gaussian 
distribution for the spectral events in this case and 
define our $\chi^2$ function as,
\be
\chi^2_{KLspec} = \sum_{i,j}(N_{i}^{exp} - N_{i}^{th})
(\sigma_{ij}^2)_{KL}^{-1}(N_{j}^{exp} - N_{j}^{th})
\ee
where $(\sigma_{ij}^2)_{KL}$ is the error correlation matrix, 
containing the statistical and systematic errors, where 
the systematic errors are assumed to be fully correlated 
among the energy bins. For the 
data we assume the same threshold of 2.6 MeV and the same 
energy  binning as the present data. 
However the current \kl data quotes a  conservative value 
of 6.42\% for 
the systematic uncertainty.  
In the 3 kton year time span 
the systematic uncertainty is expected to reduce.
The fiducial volume 
uncertainty which contributes the most at present can go down 
after the collaboration installs a calibration arm. 
The cut systematics 
and energy reconstruction error can also reduce \footnote
{We thank Prof. A.Suzuki, Prof. F. Suekane, 
Prof. S. Pakvasa and Prof. 
R. Svoboda for discussions on future systematic errors in 
KamLAND.}.  
Keeping this in mind for 3 kTy data we use a optimistic assessment
of 3\% for the 
\kl systematic error, though we have checked that increasing 
the systematic uncertainty to 4\% (which is the projected value 
quoted by the \kl collaboration) does not change the final 
conclusions. 

Figure \ref{kamspec} 
shows the reconstituted  C.L. allowed contours 
(1$\sigma$ and 3 $\sigma$)
in parameter space from  
a $\chi^2$ analysis using the simulated KamLAND spectrum alone. 
Marked by black dots are the points at which the spectra have been 
simulated.   
Each panel in this set of plots serves to identify  regions
in the presently allowed LMA region
and demonstrates the constraining capabilities of the future  
KamLAND spectral data.
Panels 1-5 are for spectrum simulated in the 
low-LMA zone while panels 
6-9 are for spectrum simulated in the high-LMA zone. 
For the first three panels the spectrum is 
simulated at a \dm corresponding 
to the low-LMA best-fit and three different $\theta$s. 
A comparison of the allowed regions in these three panels 
show that tighter constraints on $\Delta m^2$ and $\tan^2\theta$ 
are associated with
higher values of $\tan^2\theta$.
For most of the cases where \dm lies in the low-LMA region, 
the high-LMA can be ruled out at the $3\sigma$ level.
However for the panel 2 which corresponds to the 
lowest $\theta$ a small region still remains allowed at 3$\sigma$. 
The panel 4 is for the spectrum simulated at the 
best-fit solar point, $\Delta m^2=6.06\times 10^{-5}$ eV$^2$ and 
$\tan^2\theta=0.41$. 
As we move our simulation point in the high-LMA region 
in panel 6, a higher \dm region becomes allowed. 
No such region is obtained for the current best-fit
 point in the high-LMA 
region in panel 7 as well as panel 8. 
Note that panel 8 is for the high-LMA 
best-fit \dm but at an increased $\theta$. Again in panel 9 we get 
some allowed regions at 3$\sigma$ at a high \dm.  
We stress that since the relevant probability for 
\kl is the vacuum oscillation 
probability each of these panels will admit the mirror solution 
corresponding to $\theta \rightarrow \pi/2 - \theta$, the so called 
dark zone, which we have not shown explicitly in this figure. 
However we note that while very tight constraints are obtained 
for \dm, the range of allowed values for $\tan^2\theta$ seems 
to be quite large in general for all the cases. In fact for 
all but panel 2, we see that maximal mixing is allowed
atleast at the $3\sigma$ level, if not better.

Figure \ref{spectrum} shows the distorsions 
in the simulated spectrum with $\pm 1\sigma$
error bars at the
$\Delta m^2$ and $\tan^2\theta$ corresponding to each of the panels 
of figure \ref{kamspec}. 
The spectra in panels 6 and 9 of Figure 5 are simulated at a 
higher value of \dm and hence this leads to  
faster oscillations as compared to the other panels, 
resulting in a degenerate solution appearing at a higher value of 
\dm in figure \ref{kamspec}.

In figure \ref{solkamspec} we show the C.L. allowed regions from a 
combined analysis of the global solar and the 3 kTy
KamLAND simulated spectrum data. 
The KamLAND spectra are simulated 
at the same points as in Figure \ref{kamspec} and \ref{spectrum}.
Also superimposed is the 
99\% C.L.  combined solar + present KL contour.   
If we compare Figure \ref{solkamspec} to 
Figure \ref{kamspec}, we 
note that the contours get more constricted 
and the higher \dm regions appearing in panels 2, 6 and 9  
disappear. 
Inclusion of global solar data also restricts the allowed range 
in $\tan^2\theta$. 
The maximal mixing solution and the dark zone allowed by the 
\kl spectrum data get disfavoured with the inclusion of solar data 
in most of the panels, as the SNO data disfavours 
the maximal mixing solution. 
However for panels 3 and 8 for which the simulation point is at a 
larger $\theta$ the maximal mixing solution continues to be 
allowed at $3\sigma$, as
these points correspond to spectra for which the 
\kl contribution to $\chi^2$ is much less at maximal mixing.

In figure \ref{sno7} we study the effect of reducing the SNO NC 
error in the analysis with 3 kTy \kl spectrum data. 
The current SNO NC data is due to neutron capture on deuteron the 
efficiency for which is 30\%. SNO has added NaCl in the detector 
which has the effect of enhancing the neutron capture efficiency to 
83\%. With this  
the statistical error in SNO NC measurements is expected to reduce
to about 5\% \cite{snosalt}. The SNO collaboration 
are also improving their systematics. They   
expect to have less than 3\% systematic uncertainty on neutron 
capture and would also reduce the 
uncertainties in their energy scale
and energy resolution \cite{snosalt}.
Keeping this in mind we assume an optimistic reduction in the 
current systematic error of 9\% and we use
a total SNO NC error of 
7\% \cite{th12} in place of the current 12\% in figure \ref{sno7}.
We do a $\chi^2$-analysis of global solar data 
and the simulated \kl spectrum data. 
In plotting this figure, we use 
the total charged current and neutral current
rates instead of the SNO spectrum data 
which implicitly assumes absence of spectral distortion for 
the resultant $^8B$ flux at SNO.
Since we are in the LMA region we consider this to be a  
justified approximation above the SNO threshold of 5 MeV. 
This figure is a most optimistic projection of how far the allowed 
ranges of \dm and $\tan^2\theta$ can be constrained in the future. 
We find that the reduction of the SNO NC error reduces the allowed 
range of $\theta$ as compared to that in figure \ref{solkamspec}. 
In particular in the panels 3,8 and 9 maximal mixing were allowed 
with 3 kTy \kl data but reducing the SNO NC error disfavours maximal
mixing in these panels.

In Table \ref{range} we summarise the allowed ranges in
\dm and $\tan^2\theta$ 
as obtained from analysis with 
solar + 3 kTy \kl spectrum data 
and  solar (with sno NC 7\%) and 3 kTy \kl 
spectrum data. Note that for 3 kTy \kl 
spectrum we use an optimistic value of 
3\% for the \kl systematic error. 
We also give the spread \cite{th12} in each parameter where 
\be
{\rm spread} = \frac{ {\cal P}_{max} - {\cal P}_{min}}
{{\cal P}_{max} + {\cal P}_{min}}\times 100~\%
\label{error}
\ee
where ${\cal P}$ denotes the parameter \dm or $\tan^2\theta$.

We note that \kl has unprecedented sensitivity 
to \dm and the 99\% C.L. spread
reduces to 30\% 
with the 0.162 kTy \kl data. With the inclusion of 3 kTy \kl data
and improved systematics  
\dm is determined with 6\% precision. The $\theta$ 
sensitivity on the other 
hand does not improve much with the inclusion of \kl data  
The neutrino energies
corresponding to the statistically significant region of the observed
KamLAND spectra are close to the value required to make the probability
(defined in
equation \ref{prob}) one. 
Hence this region of the spectra has little sensitivity to
theta and overall the sensitivity to the value of $\theta$ is reduced.
If we reduce the SNO NC error, the large values of the 
mixing angle are severely constrained and 
the theta sensitivity improves. The probability relevant for SNO being the 
adiabatic  MSW probability it goes as $\sin^2\theta$ giving it a greater 
sensitivity to constrain mixing angles \footnote{Note however that 
in the allowed LMA region maximum $\theta$ sensitivity is obtained 
if we have a minimum in the survival probability for which 
the oscillatory term goes to 1 \cite{th12}.}.

\section{Conclusions}
We have studied  the capability of the future KamLAND data 
to constrain the mass and mixing parameters in the LMA region. 
We have simulated the \kl spectrum
in the currently allowed LMA region for an exposure of 3 kTy. 
We find that  
with 3 kTyr exposure the \kl spectrum 
data is quite powerful in constraining \dm 
giving unique allowed islands around the point at which the spectrum is 
simulated due
to it's high sensitivity to distortions driven by
this parameter. We present a few cases where the spectra is somewhat flatter 
admitting multiple solutions. However this remaining ambiguity is 
removed with the 
inclusion of the solar data in the analysis irrespective of whether the 
spectrum is simulated at points belonging to low-LMA or high-LMA zone.
Thus with 3 kton year exposure \kl can define \dm very precisely to within 
6\%. The $\theta$ sensitivity of \kl is not as good and after 3 kton year 
exposure and improved systematics the allowed spread in $\tan^2\theta$ is 
$\sim$ 37\%. 
We find that for most of the simulated spectra the maximal mixing solution is 
disfavoured but may still get allowed at 3$\sigma$ if the true spectrum corresponds to higher value of $\tan^2\theta$. 
If however the SNO NC error is reduced then the maximal mixing gets 
disfavored in all the cases. 

\begin{table}
\begin{center}
\begin{tabular}{cccccc}
\hline\hline
Data & 99\%  CL &99\% CL & 99\% CL
&  99\% CL \cr
set & range of & spread  &range  & spread
\cr
used & $\dm\times$ & of
of
& in & in \cr
 & 10$^{-5}$eV$^2$ & \dm &
$\tan^2\theta$ & $\tan^2\theta$ \cr
\hline\hline
only sol & 3.2 - 2.4
& 76\% & $.27-.75$ & 47\% \cr
sol+162 Ty &  5.3 - 9.9
& 30\% &
 $.28 -.78$ & 47\%  \cr
sol+3 KTy & 6.8 - 7.7
& 6\%  
& $.32-.70$ & 37\% \cr
sol(SNO NC 7\%)+3 KTy &  6.9 - 7.7 & 5.5\% & .33 - .60 &  29\%\cr
\hline\hline
\end{tabular}
\label{range}
\end{center}
\caption
{The range of parameter values allowed.}
\end{table}


\end{document}